\documentclass[11pt,letterpaper]{article}
\usepackage{amssymb}
\usepackage{epsfig}
\usepackage{cite}
\usepackage[headsep=0pt, head=30pt,foot=30pt,includeheadfoot,dvips,
  inner=3cm,outer=3cm,top=2cm, bottom=2cm]{geometry}

\newcommand{\labe}[1]{\label{equ:#1}}
\newcommand{\labf}[1]{\label{fig:#1}}
\newcommand{\labt}[1]{\label{tab:#1}}
\newcommand{\refe}[1]{\ref{equ:#1}}
\newcommand{\reff}[1]{\ref{fig:#1}}
\newcommand{\reft}[1]{\ref{tab:#1}}
\newcommand{\Ref}[1]{Ref.~\cite{#1}}
\newcommand{\Eq}[1]{Eq.~(\refe{#1})}
\newcommand{\Eqss}[2]{Eqs.~(\refe{#1}) and (\refe{#2})}
\newcommand{\Eqssor}[2]{Eqs.~(\refe{#1}) or (\refe{#2})}
\newcommand{\Eqsss}[3]{Eqs.~(\refe{#1}), (\refe{#2}) and (\refe{#3})}
\newcommand{\Fig}[1]{Fig.~\reff{#1}}
\newcommand{\Table}[1]{Table~\reft{#1}}

\def\collab#1{{\bf #1\rm}}
\def\etalcollab#1{\etal\,(\collab{#1}),} 

\def\Bd{$B{^0_d}$}
\def\Bdbar{$\overline B{^0_d}$}
\def\Bs{$B{^0_s}$}
\def\Bsbar{$\overline B{^0_s}$}
\def\Bqq{$B{^0_q}$}
\def\Bqbar{$\overline B{^0_q}$}
\def\BdBdbar{\Bd--\Bdbar}
\def\BsBsbar{\Bs--\Bsbar}
\def\BqBqbar{\Bqq--\Bqbar}
\def\BBbar{$B{^0}\hbox{--}\overline B{^0}$}
\def\KKbar{$K{^0}\hbox{--}\overline K{^0}$}

\def\EQN#1{\labe{#1}}
\let\DELTA=\Delta 
\def\IndexPageno#1{}
\def\lsim{\,\hbox{\char'056}\,} 
\def\etal{\hbox{\it et~al.}} 
\def\arns#1,#2(#3)
   {{\rm Ann.\,Rev.\,Nucl.\,Sci.\ }{\bf #1}, {\rm#2} {\rm(#3)}} 
\def\epjC#1,#2(#3){{\rm Eur.\,Phys.\,J.\,}{\bf C#1}, {\rm#2} {\rm(#3)}} 
\def\npB#1,#2(#3){{\rm Nucl.\,Phys.\,}{\bf B#1}, {\rm#2} {\rm(#3)}} 
\def\ptp#1,#2(#3){{\rm Prog.\,Theor.\,Phys.\,}{\bf #1}, {\rm#2} {\rm(#3)}} 
\def\plB#1,#2(#3){{\rm Phys.\,Lett.\,}{\bf B#1}, {\rm#2} {\rm(#3)}} 
\def\nim#1,#2(#3)
   {{\rm Nucl.\,Instrum.\,Methods\,}{\bf #1}, {\rm#2} {\rm(#3)}}
\def\prl#1,#2(#3){{\rm Phys.\,Rev.\,Lett.\,}{\bf #1}, {\rm#2} {\rm(#3)}} 
\def\prD#1,#2(#3){{\rm Phys.\,Rev.\,}{\bf D#1}, {\rm#2} {\rm(#3)}} 
\def\zpC#1,#2(#3){{\rm Z.\,Phys.\,}{\bf C#1}, {\rm#2} {\rm(#3)}} 
\def\ijmpA#1,#2(#3)
   {{\rm Int.\,J.\,Mod.\,Phys.\,}{\bf A#1}, {\rm#2} {\rm(#3)}}
\def\npBps#1,#2(#3){{\rm Nucl.\,Phys.\,(Proc.\,Supp.),}{\bf B#1},
{\rm#2} {\rm(#3)}} 
\def\reference#1{\bibitem{#1}}
\def\endreference{}

\def\lsim{\mathrel{\rlap{\lower4pt\hbox{\hskip1pt$\sim$}}
    \raise1pt\hbox{$<$}}}                

\begin{document}
\pagestyle{empty}
\begin{flushright}
LPHE 2008-05 \\
June 30, 2008 \\~
\end{flushright}

\vfill

\begin{center}

{\LARGE\bf \boldmath $B{^0}\hbox{--}\overline B{^0}$ MIXING}
\\[6ex] {\Large O.~SCHNEIDER} \\[1ex]
{\it Laboratoire de Physique des Hautes Energies} \\
{\it Ecole Polytechnique F\'ed\'erale de Lausanne (EPFL) \\
CH--1015 Lausanne, Switzerland} \\[1ex]
{\it e-mail:} {\tt Olivier.Schneider@epfl.ch} \\ ~

\vfill

\begin{minipage}{0.8\textwidth}
The subject of particle-antiparticle mixing in the neutral $B$ meson systems 
is reviewed. The formalism of $B{^0}\hbox{--}\overline B{^0}$ mixing 
is recalled and 
basic Standard Model predictions are given, before experimental issues are 
discussed and the latest combinations of experimental results on 
mixing parameters are presented, including those on 
mixing-induced $CP$ violation, mass differences, and decay-width differences.
Finally, time-integrated mixing results are used to improve our knowledge on 
the fractions of the various $b$-hadron species produced in $Z$ decays and 
at high-energy colliders.
\end{minipage}

\vfill ~ \vfill

{\it To appear in the 2008 edition of the 
``Review of Particle Physics'',
} \\ {\it  
C.~Amsler et al.\ (Particle Data Group), 
Phys.\ Lett.\ B (2008).
}\\[2ex]
\end{center}

\newpage
~

\newpage
\pagestyle{plain}\setcounter{page}{1}   


\begin{center}
{\LARGE\bf \boldmath \BBbar\ MIXING}

\vspace{3mm}
{\em
Updated April 2008
by O.\ Schneider (Ecole Polytechnique F\'ed\'erale de Lausanne)
}
\vspace{5mm}
\end{center}

There are two neutral \BBbar\ meson systems, \BdBdbar\ and \BsBsbar\
(generically denoted \BqBqbar, $q=s,d$), which 
exhibit particle-antiparticle mixing\cite{textbooks}.
This mixing phenomenon is described in \Ref{CP_review}.
In the following, we adopt the notation introduced in \Ref{CP_review},
and assume $CPT$ conservation throughout.
In each system, the light (L) and heavy (H) mass eigenstates,
\begin{equation} 
|B_{\rm L,H}\rangle = p | B{^0_q}\rangle \pm q |\overline B{^0_q}\rangle \,,
\EQN{eigenstates}
\end{equation} 
have a mass difference $\DELTA m_q = m_{\rm H} -m_{\rm L} > 0$,
and a total decay width difference
$\DELTA \Gamma_q = \Gamma_{\rm L} -\Gamma_{\rm H}$.
In the absence of $CP$ violation in the mixing,
$|q/p|=1$, these differences are given by $\DELTA m_q =2|M_{12}|$
and $|\DELTA \Gamma_q| =2|\Gamma_{12}|$, where $M_{12}$ and $\Gamma_{12}$
are the off-diagonal elements of the mass and decay matrices\cite{CP_review}.
The evolution of a pure $| B{^0_q}\rangle$ or
$|\overline B{^0_q}\rangle$ state at $t=0$ is given by
\begin{eqnarray} 
| B{^0_q}(t)\rangle &=& g_+(t) \,| B{^0_q}\rangle
                     + \frac{q}{p} \, g_-(t) \,|\overline B{^0_q}\rangle \,,
\EQN{time_evol1} 
\\
|\overline B{^0_q}(t)\rangle &=& g_+(t) \,|\overline B{^0_q}\rangle
                     + \frac{p}{q} g_-(t) \,| B{^0_q}\rangle \,,
\EQN{time_evol2} 
\end{eqnarray} 
which means that the flavor states remain unchanged ($+$) or oscillate
into each other ($-$) with time-dependent probabilities proportional to
\begin{equation} 
\left| g_{\pm}(t)\right|^2 = \frac{e^{-\Gamma_q t}}{2}
\left[ \cosh\!\left(
\frac{\DELTA\Gamma_q}{2}\,t\right) \pm \cos(\DELTA m_q\,t)\right] \,, 
\EQN{cosh_cos}
\end{equation} 
where $\Gamma_q = (\Gamma_{\rm H} +\Gamma_{\rm L})/2$.
In the absence of $CP$ violation, the time-integrated mixing probability
$\int \left| g_-(t)\right|^2 dt /
(\int \left| g_-(t)\right|^2 dt + \int \left| g_+(t)\right|^2 dt)$
is given by 
\begin{equation} 
\chi_q = \frac{x_q^2+y_q^2}{2(x_q^2+1)} \,, ~~~{\rm where}~~~
x_q = \frac{\DELTA m_q}{\Gamma_q}  
\,, ~~~
y_q = \frac{\DELTA \Gamma_q}{2\Gamma_q} \,.
\EQN{chi}
\end{equation} 

\section*{Standard Model predictions and phenomenology}

In the Standard Model, the transitions \Bqq$\to$\Bqbar\ and \Bqbar$\to$\Bqq\ 
are due to the weak interaction.
They are described, 
at the lowest order, by box diagrams involving
two $W$~bosons and two up-type quarks (see \Fig{box}), 
as is the case for \KKbar\ mixing.
However, the long range 
interactions arising from intermediate virtual states are negligible 
for the neutral $B$ meson systems,
because the large $B$ mass is off the region of hadronic resonances. 
The calculation of the dispersive and 
absorptive parts of the box diagrams yields the following predictions 
for the off-diagonal element of the mass and decay matrices\cite{Buras84},
\begin{figure}\begin{center}
~
\hfill
\epsfig{figure=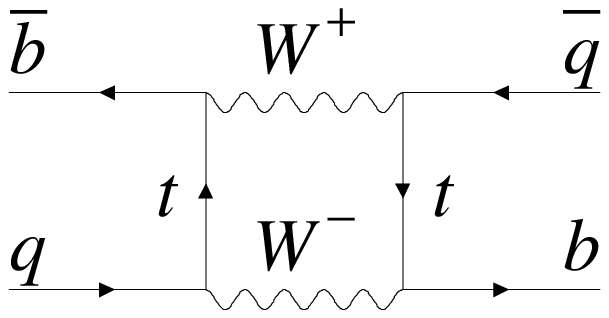,width=0.4\textwidth,clip=t}%
\hfill
\epsfig{figure=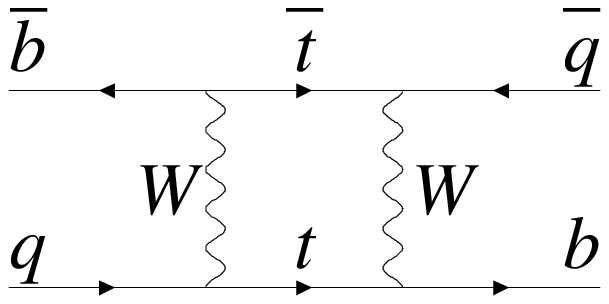,width=0.4\textwidth,clip=t}%
\hfill
~
\caption{ 
Dominant box diagrams for the 
\Bqq$\to$\Bqbar\ transitions ($q = d$ or $s$). Similar 
diagrams exist where one or both $t$ quarks are 
replaced with $c$ or $u$ quarks.
} 
\labf{box}
\end{center}\end{figure}
\begin{eqnarray} 
\hspace{-2.5ex} M_{12} \hspace{-1ex}&=&\hspace{-1ex} - \frac{
           G_F^2 m_W^2 \eta_B m_{B_q} B_{B_q} f_{B_q}^2}{12\pi^2} 
           \, S_0(m_t^2/m_W^2) \, (V_{tq}^* V_{tb}^{})^2 \,,
\EQN{M_12} \\ 
\hspace{-2.5ex} \Gamma_{12} \hspace{-1ex}&=&\hspace{-1ex} \frac{
           G_F^2 m_b^2 \eta'_B m_{B_q} B_{B_q} f_{B_q}^2}{8\pi} 
\hspace{-0.5ex}
           \left[ (V_{tq}^* V_{tb}^{})^2 + 
                   V_{tq}^* V_{tb}^{} V_{cq}^* V_{cb}^{} \,\, 
                         \hspace{-0.5ex}{\cal O}\!\left(\frac{m_c^2}{m_b^2}\right) 
                 + (V_{cq}^* V_{cb}^{})^2 \,\,
                         \hspace{-0.5ex}{\cal O}\!\left(\frac{m_c^4}{m_b^4}\right) 
           \right] \hspace{-0.5ex}\,,
\end{eqnarray} 
\noindent 
where $G_F$ is the Fermi constant, $m_W$ the $W$ boson mass,
and $m_i$ the mass of quark $i$;
$m_{B_q}$, $f_{B_q}$ and $B_{B_q}$ are the \Bqq\ mass, 
weak decay constant and bag parameter, respectively.
The known function $S_0(x_t)$ 
can be approximated very well by
$0.784\,x_t^{0.76}$\cite{Buras_Fleischer_HeavyFlavorsII},
and $V_{ij}$ are the elements of the CKM matrix\cite{CKM}.
The QCD corrections $\eta_B$ and $\eta'_B$ are of order unity. 
The only non-negligible contributions to $M_{12}$ are from box diagrams
involving two top quarks. 
The phases of $M_{12}$ and $\Gamma_{12}$ satisfy
\begin{equation} 
\phi_{M} - \phi_{\Gamma} = 
\pi + {\cal O}\left(\frac{m^2_c}{m^2_b} \right) \,,
\EQN{phasediff}
\end{equation} 
implying that the mass eigenstates  
have mass and width differences of opposite signs. This means that,
like in the $K{^0}\hbox{--}\overline K{^0}$ system, the 
heavy state is expected to have a smaller decay width than 
that of the light state: 
$\Gamma_{\rm H} < \Gamma_{\rm L}$.
Hence, $\DELTA\Gamma = \Gamma_{\rm L} -\Gamma_{\rm H}$
is expected to be positive in the Standard Model.

Furthermore, the quantity 
\begin{equation} 
\left|\frac{\Gamma_{12}}{M_{12}}\right| \simeq \frac{3\pi}{2}
\frac{m^2_b}{m^2_W} \frac{1}{S_0(m_t^2/m_W^2)} 
\sim {\cal O}\left(\frac{m^2_b}{m^2_t} \right)
\EQN{G12overM12} 
\end{equation} 
is small, and a power expansion of $|q/p|^2$ yields
\begin{equation} 
\left|\frac{q}{p}\right|^2 = 1 + \left|\frac{\Gamma_{12}}{M_{12}}\right| 
\sin(\phi_{M}-\phi_{\Gamma})
+ {\cal O}\left( \left|\frac{\Gamma_{12}}{M_{12}}\right|^2\right) \,.
\end{equation} 
Therefore, considering both 
\Eqss{phasediff}{G12overM12},
the $CP$-violating parameter
\begin{equation} 
1 - \left|\frac{q}{p}\right|^2 \simeq 
{\rm Im}\left(\frac{\Gamma_{12}}{M_{12}}\right) 
\end{equation} 
is expected to be very small: $\sim {\cal O}(10^{-3})$ for the 
\BdBdbar\ system and $\lsim {\cal O}(10^{-4})$ 
for the \BsBsbar\ system\cite{Bigi}.

In the approximation of negligible $CP$ violation in mixing, 
the ratio $\DELTA\Gamma_q/\DELTA m_q$ is equal to the 
small quantity $\left|\Gamma_{12}/M_{12}\right|$ of \Eq{G12overM12}; it is
hence independent of CKM matrix elements, {\it i.e.},
the same for the \BdBdbar\ and \BsBsbar\ systems. 
Recent calculations\cite{Lenz_Nierste} yield $\sim 5 \times 10^{-3}$ 
with a $\sim 20\%$ uncertainty.
Given the current experimental knowledge  
on the mixing parameter $x_q$ 
\begin{equation} 
\cases{
x_d = 0.776 \pm 0.008 & (\BdBdbar\ system) \cr
x_s = 26.1~ \pm 0.5~~ & (\BsBsbar\ system) \cr}
 \,,
\end{equation} 
the Standard Model thus predicts that $\DELTA\Gamma_d/\Gamma_d$ is very 
small (below 1\%), 
but $\DELTA\Gamma_s/\Gamma_s$ considerably larger 
($\sim 10\%$). These width differences are
caused by the existence of final states to which 
both the \Bqq\ and \Bqbar\ mesons can decay. Such decays involve 
$b \to c\overline{c}q$ quark-level transitions,
which are 
Cabibbo-suppressed if $q=d$ and Cabibbo-allowed if $q=s$.

A recent and complete set of Standard Model predictions for all mixing parameters 
in both the \BdBdbar\ and \BsBsbar\ systems can be found in \Ref{Lenz_Nierste}. 


\section*{Experimental issues and methods for oscillation analyses}

Time-integrated measurements of \BBbar\ mixing
were published for the first time in 1987 by UA1\cite{UA1_CP} 
and ARGUS\cite{ARGUS_CP}, 
and since then by many other experiments.
These measurements are typically based on counting same-sign and opposite-sign 
lepton pairs from the semileptonic decay of the produced $b\overline{b}$ pairs.
Such analyses cannot easily separate the contributions from the 
different $b$-hadron species, therefore, the clean environment 
of $\Upsilon(4S)$ machines (where only \Bd\ and charged $B_u$ mesons 
are produced) is in principle best suited to measure $\chi_d$.

However, better sensitivity is obtained from time-dependent analyses
aiming at the direct measurement of the oscillation frequencies 
$\DELTA m_d$ and $\DELTA m_s$,
from the proper time
distributions of \Bd\ or \Bs\ candidates 
identified through their decay in (mostly) flavor-specific modes, and
suitably tagged as mixed or unmixed.
This is particularly true for the \BsBsbar\ 
system, where the large 
value of $x_s$ implies maximal mixing, {\it i.e.}, $\chi_s \simeq 1/2$.
In such analyses, the \Bd\ or \Bs\ mesons are 
either fully reconstructed, partially reconstructed from a charm meson, 
selected from a lepton 
with the characteristics of a $b\to\ell^-$ decay,
or selected from a reconstructed displaced vertex. 
At high-energy colliders (LEP, SLC, Tevatron), 
the proper time $t=\frac{m_B}{p}L$ is measured 
from the distance $L$ between the production vertex and 
the $B$ decay vertex, 
and from an estimate of the $B$ momentum $p$.
At asymmetric $B$ factories (KEKB, PEP-II), producing 
$e^+e^-\to\Upsilon(4S) \to\hbox{\Bd\Bdbar}$ events with a boost
$\beta\gamma$ ($=0.425$, $0.55$),
the proper time difference between the two $B$ candidates
is estimated as $\DELTA t \simeq \frac{\DELTA z}{\beta\gamma c}$, 
where $\DELTA z$ is the spatial separation between the 
two $B$ decay vertices along the boost direction. 
In all cases, the good resolution needed on the vertex positions 
is obtained with silicon detectors. 

The average statistical significance ${\cal S}$ 
of a \Bd\ or \Bs\ oscillation signal can be approximated as\cite{amplitude}
\IndexPageno{Bstatsigm}
\begin{equation} 
{\cal S} \approx \sqrt{N/2} \,f_{\rm sig}\, (1-2\eta)\,
e^{-(\DELTA m\,\sigma_t)^2/2}  \,, 
\EQN{significance} 
\end{equation} 
where $N$ is the number of selected and tagged candidates, 
$f_{\rm sig}$ is the fraction of signal in that sample, 
$\eta$ is the total mistag probability, 
and $\sigma_t$ is the resolution on proper time (or proper time difference). 
The quantity ${\cal S}$ decreases very quickly as 
$\DELTA m$ increases; this dependence is controlled by $\sigma_t$,
which is therefore a critical parameter for $\DELTA m_s$ analyses. 
At high-energy colliders, the proper time resolution 
$\sigma_t \sim \frac{m_B}{\langle p\rangle} \sigma_L 
\oplus t \frac{\sigma_p}{p}$ 
includes a constant contribution due to the decay length resolution 
$\sigma_L$ (typically 0.05--0.3~ps), and a term due to the 
relative momentum resolution $\sigma_p/p$ (typically 10--20\%
for partially reconstructed decays), 
which increases with proper time.
At $B$ factories,
the boost of the $B$ mesons is estimated from the known beam energies,
and the term due to the spatial resolution dominates
(typically 1--1.5~ps because of the much smaller $B$ boost).

In order to tag a $B$ candidate 
as mixed or unmixed, it is necessary
to determine its flavor 
both in the initial state and in the final state. 
The initial and final state mistag probabilities,\IndexPageno{Bmistagm}
$\eta_i$ and $\eta_f$, degrade ${\cal S}$
by a total factor $(1-2\eta)=(1-2\eta_i)(1-2\eta_f)$.
In lepton-based analyses, the final state is tagged by the charge of 
the lepton from $b\to\ell^-$ decays; the largest contribution to $\eta_f$ 
is then due to $\overline{b}\to\overline{c}\to\ell^-$ decays. 
Alternatively, the charge of a 
reconstructed charm meson ($D^{*-}$ from \Bd\ or $D_s^-$ from \Bs), 
or that of a kaon hypothesized to come from a $b\to c\to s$
decay\cite{SLD_dmd_prelim}, can be used.
For fully inclusive analyses based on topological 
vertexing, final state tagging techniques include 
jet charge\cite{ALEPH_dmd_prelim} and charge 
dipole\cite{SLD_dms_dipole,DELPHI_dmd_dms} methods.

At high-energy colliders, the methods to tag the initial state 
({\it i.e.}, the state at production), 
can be divided into two groups: the ones 
that tag the initial charge of the $\overline{b}$ quark contained in the 
$B$ candidate itself (same-side tag),\IndexPageno{bmixam}
 and the ones that tag the initial 
charge of the other $b$ quark produced in the event (opposite-side tag). 
On the same side, the charge of a track from the primary vertex is 
correlated with the production state of the $B$ if that track is a decay
product of a $B^{**}$ state or the first particle in the fragmentation 
chain\cite{CDF_dmd,ALEPH_dms}.
Jet- and vertex-charge techniques work on both sides and on the opposite
side, respectively. 
Finally, the charge of a lepton from $b\to\ell^-$ or of a kaon
from $b\to c\to s$ can be used as opposite side tags,
keeping in mind that their performance is degraded due to integrated mixing.
At SLC, the beam polarization produced a sizeable forward-backward 
asymmetry in the $Z\to b\overline{b}$ decays, and provided another
very interesting and effective initial state tag based on the polar angle 
of the $B$ candidate\cite{SLD_dms_dipole}.
Initial state tags have also been combined to 
reach $\eta_i \sim 26\%$ at LEP\cite{ALEPH_dms,DELPHI_dms_dgs},
or even 22\% at SLD\cite{SLD_dms_dipole} with full efficiency. 
In the case $\eta_f=0$, this corresponds to an effective tagging efficiency
$Q=\epsilon D^2=\epsilon(1-2\eta)^2$, 
where $\epsilon$ is the tagging efficiency,
in the range $23-31\%$. 
The equivalent figure achieved by CDF 
during Tevatron Run~I was $\sim3.5\%$\cite{Paulini}
reflecting the fact that tagging is more difficult at hadron colliders.
The current CDF and D\O\ analyses of Tevatron Run~II data reach 
$\epsilon D^2 = (1.8\pm0.1)\%$\cite{CDF2_dms} 
and $(2.5\pm0.2)\%$\cite{DZERO_dmd} for opposite-side tagging, 
while same-side kaon tagging (for \Bs\ oscillation analyses) 
is contributing an additional $3.7-4.8\%$ 
at CDF\cite{CDF2_dms} and pushes the combined performance to 
$(4.5\pm0.9)\%$ at D\O\cite{DZERO_dms_prelim}.

At $B$ factories, the flavor of a \Bd\ meson at production cannot 
be determined, since the two neutral $B$ mesons produced in a 
$\Upsilon(4S)$ decay evolve in a coherent $P$-wave state where they 
keep opposite flavors at any time.
However, as soon as one of them decays, the other follows a time-evolution
given by \Eqssor{time_evol1}{time_evol2}, 
where $t$ is replaced with $\DELTA t$
(which will take negative values half of the time).
Hence, the ``initial state''
tag of a $B$ can be taken as the final state tag of the other $B$. 
Effective tagging efficiencies $Q$ of 30\%
are achieved by BABAR 
and Belle\cite{BABAR_Belle_tagging}, 
using different techniques including $b \to \ell^-$ and 
$b\to c\to s$ tags. 
It is worth noting that, in this case, 
mixing of the other $B$ ({\it i.e.}, the coherent mixing occurring before
the first $B$ decay) does not contribute to the mistag probability.

In the absence of experimental observation of a decay-width difference,
oscillation analyses typically neglect $\DELTA\Gamma$ in \Eq{cosh_cos},
and describe the data with the physics functions
$\Gamma e^{-\Gamma t} (1 \pm \cos( \DELTA m t))/2$
(high-energy colliders) or
$\Gamma e^{-\Gamma |\DELTA t|} (1 \pm \cos( \DELTA m \DELTA t))/4$
(asymmetric $\Upsilon(4S)$ machines).
As can be seen from \Eq{cosh_cos}, a non-zero value of $\DELTA\Gamma$
would effectively reduce the oscillation amplitude with a small 
time-dependent factor that would be very difficult to distinguish 
from time resolution effects. 
Measurements of $\DELTA m_d$ are usually extracted from the data
using a maximum likelihood fit.
To extract information useful for the interpretation of \Bs\ oscillation searches 
and for the combination of their results, 
a method\cite{amplitude} is followed
in which a \Bs\ oscillation amplitude ${\cal A}$
is measured as a function of a fixed test value of $\DELTA m_s$, 
using a maximum likelihood fit based on the functions
$\Gamma_s e^{-\Gamma_s t} (1 \pm {\cal A} \cos( \DELTA m_s t))/2$. 
To a good approximation, the statistical uncertainty on ${\cal A}$
is Gaussian and equal to $1/{\cal S}$ from \Eq{significance}.
If $\DELTA m_s$ is equal to its true value, one expects
${\cal A} = 1 $ within the total uncertainty $\sigma_{\cal A}$;
in case a signal is seen, its observed (or expected) 
significance will be defined as 
${\cal A}/\sigma_{\cal A}$ (or $1/\sigma_{\cal A}$).
However, if $\DELTA m_s$ is (far) below its
true value, a measurement consistent with ${\cal A} = 0$ is expected.
A value of $\DELTA m_s$ can be excluded at 95\%~CL
if ${\cal A} + 1.645\,\sigma_{\cal A} \le 1$ (since the integral
of a normal distribution from $-\infty$ to $1.645$ is equal to 0.95). 
Because of the proper time resolution, the quantity $\sigma_{\cal A}(\DELTA m_s)$
is a steadily increasing function of $\DELTA m_s$. 
We define the sensitivity for 95\%~CL exclusion of $\DELTA m_s$ values
(or for a $3\,\sigma$ or $5\,\sigma$ observation of \Bs\ oscillations) 
as the value of $\DELTA m_s$ for which 
$1/\sigma_{\cal A}=1.645$ (or $1/\sigma_{\cal A}=3$ or $5$).


\section*{\boldmath \Bd\ mixing studies}


Many \BdBdbar\ oscillations analyses have been 
published\cite{WARNING} by the ALEPH\cite{ALEPH_dmd}, 
BABAR\cite{BABAR_dmd}, Belle\cite{Belle_dmd},
CDF\cite{CDF_dmd}, D\O\cite{DZERO_dmd},
DELPHI\cite{DELPHI_dmd_dms,DELPHI_dmd}, 
L3\cite{L3_dmd}, and OPAL\cite{OPAL_dmd,OPAL_CP_semi} collaborations.
Although a variety of different techniques have been used, the 
individual $\DELTA m_d$ 
results obtained at high-energy colliders have remarkably similar precision.
Their average is compatible with the  recent and more precise measurements 
from asymmetric $B$ factories.
The systematic uncertainties are not negligible; 
they are often dominated by sample composition, mistag probability,
or $b$-hadron lifetime contributions.
Before being combined, the measurements are adjusted on the basis of a 
common set of input values, including the $b$-hadron lifetimes and fractions
published in this {\it Review}. Some measurements are statistically correlated. 
Systematic correlations arise both from common physics sources (fragmentation 
fractions, lifetimes, branching ratios of $b$~hadrons), and from purely 
experimental or algorithmic effects (efficiency, resolution, tagging, 
background description). Combining all published measurements%
\cite{DELPHI_dmd_dms,CDF_dmd,DZERO_dmd,ALEPH_dmd,BABAR_dmd,Belle_dmd,DELPHI_dmd,L3_dmd,OPAL_dmd,OPAL_CP_semi}
and accounting for all identified correlations 
yields 
$\DELTA m_d = {\rm 0.507 \pm 0.003 (stat) \pm 0.003 (syst)}~\hbox{ps}^{-1}$\cite{HFAG}, 
a result dominated by the $B$ factories.

On the other hand, ARGUS and CLEO have published time-integrated 
measurements\cite{ARGUS_chid,CLEO_chid_CP,CLEO_chid_CP_y}, 
which average to $\chi_d = 0.182 \pm 0.015$.
Following \Ref{CLEO_chid_CP_y}, 
the width difference $\DELTA \Gamma_d$ could 
in principle be extracted from the
measured value of $\Gamma_d$ and the above averages for 
$\DELTA m_d$ and $\chi_d$ 
(see \Eq{chi}),
provided that $\DELTA \Gamma_d$ has a negligible impact on 
the $\DELTA m_d$ measurements.
However, direct time-dependent studies published  
by DELPHI\cite{DELPHI_dmd_dms} and 
BABAR\cite{BABAR_DGd_qp} provide stronger constraints, which can be combined 
to yield 
$\rm{sign}(\rm{Re} \lambda_{CP}) \DELTA \Gamma_d/\Gamma_d 
= 0.009 \pm 0.037$\cite{HFAG}.

Assuming $\DELTA \Gamma_d =0$ and no $CP$ violation in mixing,
and using the measured \Bd\ lifetime of $1.530\pm0.009~{\rm ps}$,
the $\DELTA m_d$ and $\chi_d$ results are combined to yield the 
world average
\begin{equation} 
\DELTA m_d = 0.507 \pm 0.005~\hbox{ps}^{-1} 
\EQN{dmdw}
\end{equation} 
or, equivalently,
\begin{equation} 
\chi_d=0.1878\pm 0.0024\,.  
\EQN{chidw}
\end{equation} 

Evidence for $CP$ violation in \Bd\ mixing has been searched for,
both with flavor-specific and inclusive \Bd\ decays, 
in samples where the initial 
flavor state is tagged. In the case of semileptonic 
(or other flavor-specific) decays, 
where the final state tag is 
also available, the following asymmetry\cite{CP_review}
\begin{equation} 
 {\cal A}_{\rm SL}^d = 
\frac{
N(\hbox{\Bdbar}(t) \to \ell^+           \nu_{\ell} X) -
N(\hbox{\Bd}(t)    \to \ell^- \overline{\nu}_{\ell} X) }{
N(\hbox{\Bdbar}(t) \to \ell^+           \nu_{\ell} X) +
N(\hbox{\Bd}(t)    \to \ell^- \overline{\nu}_{\ell} X) } 
\simeq 1 - |q/p|^2_d 
\end{equation} 
has been measured, either in time-integrated analyses at 
CLEO
\cite{CLEO_chid_CP_y,CLEO_CP_semi},
CDF\cite{CDF_CP_semi,CDF2_CP_semi_prel} and D\O\cite{DZERO_Asl},
or in time-dependent analyses at 
LEP\cite{OPAL_CP_semi,DELPHI_CP,ALEPH_CP},
BABAR\cite{BABAR_DGd_qp,BABAR_CP_semi,BABAR_CP_semi_dstarlnu_prel} and Belle\cite{Belle_Asl}.
In the inclusive case, also investigated at 
LEP\cite{DELPHI_CP,ALEPH_CP,OPAL_CP_incl},
no final state tag is used, and the asymmetry\cite{incl_asym}
\begin{equation} 
\frac{
N(\hbox{\Bd}(t) \to {\rm all}) -
N(\hbox{\Bdbar}(t) \to {\rm all}) }{
N(\hbox{\Bd}(t) \to {\rm all}) +
N(\hbox{\Bdbar}(t) \to {\rm all}) } 
\simeq
{\cal A}_{\rm SL}^d \left[ \frac{x_d}{2} \sin(\DELTA m_d \,t) - 
\sin^2\left(\frac{\DELTA m_d \,t}{2}\right)\right] 
\end{equation} 
must be measured as a function of the proper time to extract information 
on $CP$ violation.
In all cases, asymmetries compatible with zero have been found,  
with a precision limited by the available statistics. A simple 
average of all published 
results for the \Bd\ meson%
\cite{OPAL_CP_semi,CLEO_chid_CP_y,BABAR_DGd_qp,CLEO_CP_semi,DZERO_Asl,ALEPH_CP,BABAR_CP_semi,Belle_Asl,OPAL_CP_incl}
yields
${\cal A}_{\rm SL}^d = {\rm -0.0049 \pm 0.0038}$, 
under the assumption of no $CP$ violation in \Bs\ mixing.
Published results at $B$ factories only%
\cite{CLEO_chid_CP_y,BABAR_DGd_qp,CLEO_CP_semi,BABAR_CP_semi,Belle_Asl}, 
where no \Bs\ is produced, average to
\begin{equation} 
{\cal A}_{\rm SL}^d = {\rm -0.0005 \pm 0.0056}\,, 
~ \hbox{or} ~ |q/p|_d = 1.0002 \pm 0.0028\,, 
\EQN{ASLd}
\end{equation} 
a result which does not yet constrain the Standard Model.

The $\DELTA m_d$ result of \Eq{dmdw} provides an estimate of $2|M_{12}|$, 
and can be used, 
together with \Eq{M_12}, 
to extract the magnitude of the CKM matrix element $V_{td}$ 
within the Standard Model\cite{CKM_review}. 
%
%
%
%
%
%
%
%
The main experimental 
uncertainties on the resulting estimate of $|V_{td}|$ come from 
$m_t$ and $\DELTA m_d$; however, the extraction is at present 
completely dominated by the uncertainty on the hadronic 
matrix element $f_{B_d} \sqrt{B_{B_d}} = 244\pm26$~MeV 
obtained from lattice QCD calculations\cite{lattice_QCD}.


\section*{\boldmath \Bs\ mixing studies}

In the decade before the Tevatron Run~II results became available, 
\BsBsbar\ oscillations have been the subject of many 
studies from ALEPH\cite{ALEPH_dms_final}, CDF\cite{CDF1_dms},
DELPHI\cite{DELPHI_dmd_dms,DELPHI_dms_dgs,DELPHI_dms}, 
OPAL\cite{OPAL_dms} and 
SLD\cite{SLD_dms_dipole,SLD_dms_ds,SLD_dms_prelim}.
The most sensitive analyses appeared to be the ones based 
on inclusive lepton samples. Because of their better 
proper time resolution, the small data samples analyzed 
inclusively at SLD, as well as the fully reconstructed $B_s$ decays 
at LEP were also very useful to explore the high $\DELTA m_s$ region.
However, all results were limited by the available statistics.
All published measurements of the \Bs\ oscillation amplitude%
\cite{SLD_dms_dipole,DELPHI_dmd_dms,DELPHI_dms_dgs,ALEPH_dms_final,CDF1_dms,DELPHI_dms,OPAL_dms,SLD_dms_ds}
are averaged\cite{HFAG} to yield the combined amplitudes 
${\cal A}$ shown in \Fig{amplitude} (bottom) as a function of $\DELTA m_s$. 
The individual results 
have been adjusted to common physics inputs, and all known correlations 
have been accounted for; 
the sensitivities of the inclusive analyses, 
which depend directly through \Eq{significance} 
on the assumed fraction $f_s$
of \Bs\ mesons in an unbiased sample of weakly-decaying $b$~hadrons, 
have also been rescaled to a common
average of 
$f_s = 0.104 \pm 0.009$. 
The combined sensitivity for 95\%~CL exclusion of $\DELTA m_s$ values is found to be
       18.3~ps$^{-1}$. 
All values of $\DELTA m_s$ below 
       14.6~ps$^{-1}$ 
       are excluded at 95\%~CL, while
the values between 
       14.6 and 21.7~ps$^{-1}$ 
cannot be excluded, because 
the data is compatible with a signal in this region. However,
the largest deviation from ${\cal A}=0$ in this range is a 1.9\,$\sigma$ effect
only, so no signal can be claimed.

\begin{figure}
\vspace{-25mm}
\begin{center}
\epsfig{figure=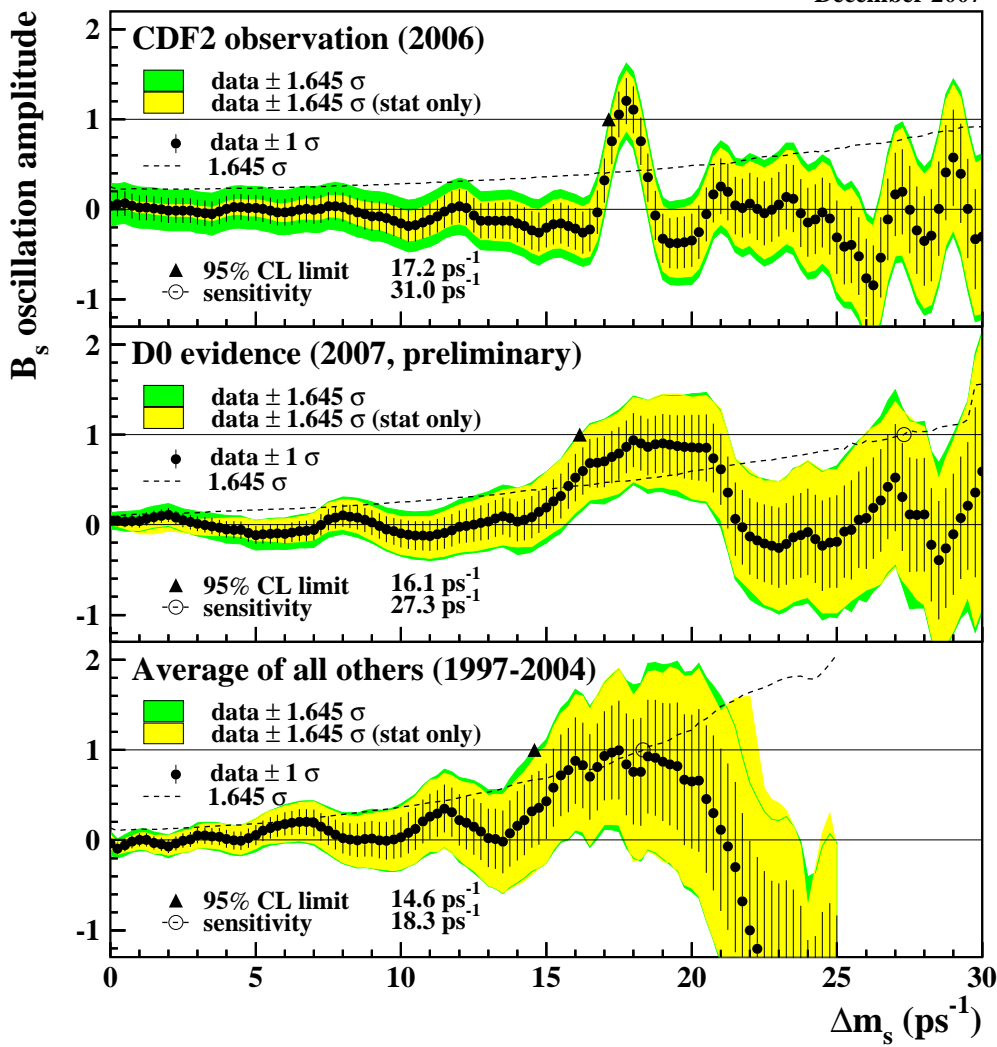,width=0.85\textwidth}
\caption{ 
Combined measurements of the \Bs\ oscillation amplitude as a function of $\DELTA m_s$.
Top: CDF result based on Run II data, published in 2006\cite{CDF2_dms}.
Middle: Average of all preliminary D\O\ results available at the end 
of 2007\cite{DZERO_dms_prelim}.
Bottom: Average of all other results (mainly from LEP and SLD) 
published between 1997 and 2004. 
All measurements are dominated by statistical uncertainties. 
Neighboring points are statistically correlated.
}
\labf{amplitude}
\end{center}\end{figure}

Tevatron Run~II results based on 1~fb$^{-1}$ of data became available in 2006. 
After D\O\cite{DZERO_dms} reported $17 < \DELTA m_s < 21~{\rm ps}^{-1}$ (90\%~CL) 
and a most probable value of 19~ps$^{-1}$ with an observed (expected) 
significance of 2.5\,$\sigma$ (0.9\,$\sigma$), CDF\cite{CDF2_dms} published 
the first direct evidence of \Bs\ oscillations shortly followed by a $>5\sigma$ observation 
(shown at the top of \Fig{amplitude}). The measured value of $\DELTA m_s$ is
\begin{equation} 
\DELTA m_s = 
{\rm 17.77 \pm 0.10 (stat) \pm 0.07 (syst)}~\hbox{ps}^{-1} \,, \EQN{CDFdms} 
\end{equation} 
based on samples of flavour-tagged
hadronic and semileptonic \Bs\ decays, partially or 
fully reconstructed in flavour-specific final states. 
More recently, D\O\cite{DZERO_dms_prelim} obtained with 2.4~fb$^{-1}$
an independent $2.9\sigma$ preliminary evidence for \Bs\ oscillations 
(middle of \Fig{amplitude}) at 
$\DELTA m_s = {\rm 18.53 \pm 0.93 (stat) \pm 0.30 (syst)}~\hbox{ps}^{-1}$\cite{DZERO_dms_combination}, 
consistent with the CDF measurement.

The information on $|V_{ts}|$ obtained
in the framework of the Standard Model 
is hampered by the hadronic uncertainty, as in the \Bd\ case. 
However, several uncertainties cancel in the frequency ratio
\begin{equation} 
\frac{\DELTA m_s}{\DELTA m_d} = \frac{m_{B_s}}{m_{B_d}}\, \xi^2
                    \left|\frac{V_{ts}}{V_{td}}\right|^2 \,,
\EQN{ratio} 
\end{equation} 
where $\xi= (f_{B_s} \sqrt{B_{B_s}})/(f_{B_d} \sqrt{B_{B_d}})
=1.210\, ^{+0.047}_{-0.035}$
is an SU(3) flavor-symmetry breaking factor 
obtained from lattice QCD calculations\cite{lattice_QCD}.
Using the measurements of 
\Eqss{dmdw}{CDFdms}, one can extract
\begin{equation} 
\left|\frac{V_{td}}{V_{ts}}\right| = 
{\rm 0.2060 \pm 0.0012 (exp)\,^{+0.0080}_{-0.0060} (lattice)} \,, \EQN{VtdVts}
\end{equation} 
in good agreement with (but much more precise than) 
the recent results obtained by the Belle\cite{Belle-btodgamma} 
and BABAR\cite{BABAR-btodgamma} collaborations based on the 
observation of the $b\to d \gamma$ transition.
The CKM matrix can be constrained using experimental results 
on observables such as $\DELTA m_d$, 
$\DELTA m_s$, $|V_{ub}/V_{cb}|$, $\epsilon_K$, and $\sin(2\beta)$ 
together with theoretical inputs and unitarity 
conditions\cite{CKM_review,UTfit,CKMfitter}.
The constraint from our knowledge on the ratio $\DELTA m_s/\DELTA m_d$
is presently more effective in limiting the position of the apex of the 
CKM unitarity triangle than the one obtained from the $\DELTA m_d$ 
measurements alone, due to the reduced hadronic uncertainty in \Eq{ratio}.
We also note that the measured value of $\Delta m_s$ is consistent with the Standard 
Model prediction obtained from CKM fits
where no experimental information on $\Delta m_s$ is used, 
{\it e.g.} $20.6 \pm 2.6~{\rm ps}^{-1}$\cite{UTfit} or
$17.7\, ^{+6.4}_{-2.1}~{\rm ps}^{-1}$\cite{CKMfitter}.

Information on $\DELTA\Gamma_s$ can be obtained by studying the proper time 
distribution of untagged 
\Bs\ samples%
\cite{Hartkorn_Moser}.
In the case of an inclusive \Bs\ selection\cite{L3_DGs}, or a semileptonic (or flavour-specific)
\Bs\ decay selection\cite{DELPHI_dms_dgs,ALEPH_OPAL_CDF_Dsl_lifetime,CDF2DZERO_tauBsfs}, 
both the short- and long-lived
components are present, and the proper time distribution is a superposition 
of two exponentials with decay constants
$\Gamma_{\rm L,H} = \Gamma_s\pm \DELTA\Gamma_s/2$.
In principle, this provides sensitivity to both $\Gamma_s$ and 
$(\DELTA\Gamma_s/\Gamma_s)^2$. Ignoring $\DELTA\Gamma_s$ and fitting for 
a single exponential leads to an estimate of $\Gamma_s$ with a 
relative bias proportional to $(\DELTA\Gamma_s/\Gamma_s)^2$. 
An alternative approach, which is directly sensitive to first order in 
$\DELTA\Gamma_s/\Gamma_s$, 
is to determine the lifetime of \Bs\ candidates decaying to $CP$ 
eigenstates; measurements exist for
$\hbox{\Bs} \to K^+K^-$\cite{CDF_tauKK},
$\hbox{\Bs} \to J/\psi \phi$\cite{CDF1_Jpsiphi,CDF2DZERO_Jpsiphi}
and $\hbox{\Bs} \to D_s^{(*)+} D_s^{(*)-}$\cite{ALEPH_DGs}, which are 
mostly $CP$-even states\cite{Aleksan}. 
However, in the case of $\hbox{\Bs} \to J/\psi \phi$ 
this technique has now been replaced 
by more sensitive time-dependent angular analyses 
that allow the simultaneous extraction of $\DELTA\Gamma_s/\Gamma_s$ and 
the $CP$-even and $CP$-odd amplitudes\cite{CDF2_Jpsiphi_angular,DZERO_Jpsiphi_angular}.
Estimates of $\DELTA\Gamma_s/\Gamma_s$
have also been obtained directly from measurements of the 
$\hbox{\Bs} \to D_s^{(*)+} D_s^{(*)-}$ branching ratio\cite{ALEPH_DGs,Bs_DsDs_other}, 
under the assumption that 
these decays account for all the $CP$-even final states 
(however, no systematic uncertainty due to this assumption is given, so 
the averages quoted below will not include these estimates).

Applying the combination procedure of \Ref{HFAG} 
(including the constraint from the flavour-specific lifetime measurements) 
on the published results\cite{DELPHI_dms_dgs,ALEPH_OPAL_CDF_Dsl_lifetime,CDF1_Jpsiphi,ALEPH_DGs,CDF2_Jpsiphi_angular,DZERO_Jpsiphi_angular}
yields
\begin{equation} 
\DELTA\Gamma_s/\Gamma_s = +0.069\, ^{+0.058}_{-0.062} 
~~~~\hbox{and}~~~
1/\Gamma_s = 1.470\, ^{+0.026}_{-0.027}~\hbox{ps}\,,
\EQN{DGs}
\end{equation} 
or equivalently
\begin{equation} 
1/\Gamma_{\rm L} = 1.419\, ^{+0.039}_{-0.038}~\hbox{ps}
~~~~\hbox{and}~~~
1/\Gamma_{\rm H} = 1.525\, ^{+0.062}_{-0.063}~\hbox{ps}\,,
\end{equation} 
under the assumption of no $CP$ violation in \Bs\ mixing. 

Recent studies also consider $CP$ violation, 
either in untagged\cite{CDF2_Jpsiphi_angular,DZERO_Jpsiphi_angular} or 
tagged\cite{CDF2_Jpsiphi_tagged,DZERO_Jpsiphi_tagged} $\hbox{\Bs} \to J/\psi \phi$ decays,
and start to constrain the phase difference $-2\beta_s$
between the \Bs\ mixing diagram and 
the $b \to c\bar{c}s$ tree decay diagram. In the Standard Model, 
$\beta_s = \arg(-(V_{ts}^{}V_{tb}^*)/(V_{cs}^{}V_{cb}^*))$ 
is expected to be about one degree\cite{Lenz_Nierste}.
A proper combination of the current Tevatron constraints on $\beta_s$ 
requires extra information not available in the original publications  
and is being prepared in collaboration between CDF and D\O.

On the other hand $CP$ violation in \Bs\ mixing
has been investigated through the asymmetry between 
positive and negative same-sign muon pairs from semi-leptonic 
decays of $b\bar{b}$ pairs\cite{CDF_CP_semi,CDF2_CP_semi_prel,DZERO_Asl}
and directly through the charge asymmetry of
$\hbox{\Bs} \to D_s \mu\nu X$ decays\cite{DZERO_Asl_Bs}.
Combining all published results\cite{CDF_CP_semi,DZERO_Asl,DZERO_Asl_Bs} with the knowledge of $CP$ violation in 
\Bd\ mixing from \Eq{ASLd} leads to 
\begin{equation} 
{\cal A}_{\rm SL}^s = -0.0030 \pm 0.0101 \,, 
~ \hbox{or} ~ |q/p|_s = 1.0015 \pm 0.0051 \,. 
\end{equation} 

A large New Physics phase 
could possibly contribute to 
both $CP$ violation in $\hbox{\Bs} \to J/\psi \phi$ and to the 
mixing phase difference of \Eq{phasediff} on which ${\cal A}_{\rm SL}^s$ depends. 
Combined fits\cite{D0_combined_phis,UTfit_NewPhysics} of $\beta_s$ and 
${\cal A}_{\rm SL}^s$ measurements already yield interesting constraints 
on this New Physics phase. 
A deviation from the Standard Model,
with a significance of more than $3\sigma$,
has recently been claimed\cite{UTfit_NewPhysics},
based on a preliminary analysis including the latest Tevatron \Bs\ mixing
results\cite{CDF2_dms,CDF2_CP_semi_prel,DZERO_Asl,CDF2_Jpsiphi_tagged,DZERO_Jpsiphi_tagged,DZERO_Asl_Bs}.
                  


\section*{\boldmath Average $b$-hadron mixing probability and 
$b$-hadron production fractions in $Z$ decays and at high energy}

Mixing measurements can 
significantly improve our knowledge on the fractions 
$f_u$, $f_d$, $f_s$ and $f_{\rm baryon}$, defined as 
the fractions of $B_u$, \Bd, \Bs\, and $b$-baryon 
in an unbiased sample of weakly decaying $b$~hadrons
produced in high-energy collisions. Indeed, 
time-integrated mixing analyses performed with lepton pairs 
from $b\overline{b}$ events at high energy 
measure the quantity 
\begin{equation} 
\overline{\chi} = f'_d \,\chi_d + f'_s \,\chi_s \,,
\end{equation} 
where $f'_d$ and $f'_s$ are the fractions of \Bd\ and \Bs\ hadrons 
in a sample of semileptonic $b$-hadron decays. 
Assuming that all $b$~hadrons have the same semileptonic 
decay width implies 
$f'_q = f_q/(\Gamma_q \tau_b)$ ($q=s,d$), where
$\tau_b$ is the average $b$-hadron lifetime. 
Hence $\overline{\chi}$ measurements, together with 
the $\chi_d$ average of \Eq{chidw} and 
the very good approximation $\chi_s = 1/2$ 
(in fact $\chi_s = 0.49927 \pm 0.00003$ from 
\Eqsss{chi}{CDFdms}{DGs}),
provide constraints on the fractions $f_d$ and $f_s$.

The LEP experiments have measured 
$f_s \times {\rm BR}(B^0_s \to D_s^- \ell^+ \nu_\ell X)$\cite{LEP_fs}, 
${\rm BR}(b \to \Lambda_b^0) \times 
{\rm BR}(\Lambda_b^0 \to \Lambda_c^+\ell^- \overline\nu_\ell X)$\cite{LEP_fla},
and ${\rm BR}(b \to \Xi_b^-) \times 
{\rm BR}(\Xi_b^- \to \Xi^-\ell^-\overline\nu_\ell X)$\cite{LEP_fxi}
from partially reconstructed final states
including a lepton, $f_{\rm baryon}$ 
from protons identified in $b$ events\cite{ALEPH-fbar}, and the 
production rate of charged $b$ hadrons\cite{DELPHI-fch}. 
The $b$-hadron fractions measured at CDF 
using double semileptonic $K^*\mu\mu$ and $\phi\mu\mu$ final states\cite{CDF_f_dsl} 
and electron-charm final states\cite{CDF_f_ec} 
are at slight discrepancy with the ones measured at LEP.
Furthermore the averages of the $\overline{\chi}$ values measured at LEP, 
$0.1259 \pm0.0042$\cite{LEPEWWG_chibar}, and at Tevatron, 
$0.147 \pm0.011$\cite{DZERO_Asl,CDF_chibar}, 
show a 1.8\,$\sigma$ deviation with respect to each other. 
This may be a hint that the fractions at the Tevatron might be different 
from the ones in $Z$ decays.
Combining\cite{HFAG} all the available information under the constraints
$f_u = f_d$ and $f_u + f_d + f_s + f_{\rm baryon} = 1$
yields the two set of averages shown in \Table{fractions}. The second set, 
obtained using both LEP and Tevatron results, has larger errors than the first set, 
obtained using LEP results only, because we have applied scale factors 
as advocated by the PDG for the treatment of marginally consistent data. 

\begin{table}
\caption{
$\overline{\chi}$ and $b$-hadron fractions (see text).
\labt{fractions}
}
\begin{center}
\begin{tabular}{lll}
\hline\hline
                     & in $Z$ decays   & at high energy   \\
\hline
$\overline{\chi}$    & $0.1259  \pm 0.0042$ ~ & $0.1284  \pm 0.0069$ \\
$f_u = f_d$          & $0.402\,~ \pm 0.009$   & $0.399\,~ \pm 0.011$ \\
$f_s$                & $0.104\,~ \pm 0.009$   & $0.110\,~ \pm 0.012$ \\
$f_{\rm baryon}$ ~~  & $0.091\,~ \pm 0.015$   & $0.092\,~ \pm 0.019$  \\
\hline\hline
\end{tabular}
\end{center}
\end{table} 



\section*{Summary and prospects}

\BBbar\ mixing has been and still is a field of intense study. While 
fairly little experimental progress was recently achieved in the \Bd\ sector,
impressive new \Bs\ results are becoming available from Run~II of the Tevatron. 
\Bs\ oscillations are now established and the 
mass difference in the \BsBsbar\ system is measured very accurately, 
with a central value compatible with the Standard Model (SM) expectation
and a relative precision ($0.7\%$) matching that in the \BdBdbar\ system ($0.9\%$). 
However, the extraction of $|V_{td}/V_{ts}|$ from these measurements 
in the SM framework is limited by the hadronic uncertainty, 
which will be an important challenge to reduce in the future.
New time-dependent angular analyses of 
$\hbox{\Bs} \to J/\psi \phi$ decays and measurements of time-integrated 
\Bs\ asymmetries at CDF and D\O\
are improving our knowledge of the other \Bs\ mixing parameters:
while $CP$ violation in \BsBsbar\ mixing 
is consistent with zero, 
with an uncertainty still large compared to the SM prediction, 
the relative decay width difference $\DELTA \Gamma_s/\Gamma_s$ is now determined
to an absolute precision of $\sim 6\%$, smaller than the central value
of the SM prediction. The data prefer
$\Gamma_{\rm L} > \Gamma_{\rm H}$ as predicted in the SM.

Improved \Bs\ results are still to come, 
with very promising short-term prospects, 
both for $\DELTA \Gamma_s$ and $CP$-violating phases induced by mixing
such as $\beta_s$ and $\arg(-M_{12}/\Gamma_{12})$. 
Although first interesting experimental constraints have been published, 
very little is known yet about these phases, which are predicted to be very small in the SM. 
A full search for New Physics effects in these observables will require 
statistics beyond that of the Tevatron. 
These will eventually become available 
at CERN's Large Hadron Collider, scheduled to start operation in 2008, where 
%
LHCb 
%
expects to be able to measure
$\beta_s$ down to the SM value after many years of operations\cite{LHCb}.

$B$ mixing may still reveal a surprize, 
but much effort is needed for this, both on the experimental and theoretical sides,  
in particular to further reduce the hadronic uncertainties of lattice QCD calculations.
In the long term, a stringent check of the consistency
of the \Bd\ and \Bs\ mixing amplitudes (magnitudes and phases) 
with all other measured flavour-physics observables
will be possible within the SM, 
leading to very tight limits (or otherwise new interesting knowledge\,!) on New Physics.

\end{document}